\documentclass[twocolumn,preprintnumbers,amsmath,amssymb]{revtex4}
\usepackage{graphicx}
\usepackage{dcolumn}
\usepackage{bm}
\usepackage{latexsym}
\begin{document}
\title{TASEP on parallel tracks: effects of mobile bottlenecks in fixed segments}
\author{Sumit Sinha and Debashish Chowdhury{\footnote{Corresponding author: debch@iitk.ac.in}}}
\affiliation{\textit{Department of Physics, Indian Institute of Technology,Kanpur 208016,India}}
\begin{abstract}
We study the flux of totally asymmetric simple exclusion processes (TASEPs) on a twin co-axial square tracks. In this biologically motivated model the particles in each track act as mobile bottlenecks against the movement of the particles in the other although the particle are not allowed to move out of their respective tracks. So far as the outer track is concerned, the particles on the inner track act as bottlenecks only over a set of fixed segments of the outer track, 
in contrast to site-associated and particle-associated quenched randomness in the earlier models of disordered TASEP. 
In a special limiting situation the movement of particles in the outer track mimic a TASEP with a ``point-like'' immobile (i.e., quenched) defect where phase segregation of the particles is known to take place. The length of the inner track as well as the strength and number density of the mobile bottlenecks moving on it are the control parameters that determine the nature of spatio-temporal organization of particles on the outer track. Variation of these control parameters allow variation of the width of the phase-coexistence region on the flux-density plane of the outer track. Some of these phenomena are likely to survive even in the future extensions intended for studying traffic-like collective phenomena of polymerase motors on double-stranded DNA.
\end{abstract}
\maketitle 

\section{Introduction} 

Totally asymmetric simple exclusion process (TASEP) is one of the simplest models of 
non-equilibrium systems of interacting self-driven particles \cite{schuetz00}. 
Properties of TASEP and its various extensions have been analyzed to get insight into 
the spatio-temporal organization in wide varieties 
of physical and biological systems \cite{chowdhury00,schadschneider10,chowdhury05,chou11,chowdhury13}. 
In the simplest version of this model particles hop forward, with rate $p$, from one 
site to the next on a one-dimensional lattice of equi-spaced sites; however, a particle 
successfully executes the forward hop if, and only if, the target site is empty. 

Effects of two types of quenched (time-independent) defects on the spatio-temporal 
organization of the particles have been explored extensively \cite{chowdhury00,schadschneider10}. 
(A) In one class of models of quenched defect the randomness is associated with the {\it lattice}: 
particles hop at the rate $p$ from all sites except from the ``point defect'' from where 
the hopping rate is $p' < p$. Such a single defect site can give rise a nontrivial 
macroscopic phase segregation of the particles into high-density and low-density regions 
\cite{janowski94}. Two types of extensions of this model have been reported: (i) $r$ ($>1$) 
successive sites are occupied by defects so that the defect may be regarded as a single  
extended object of length $r$; (ii) $r$ ($>1$) point defects are distributed randomly over 
the entire lattice \cite{tripathy97,harris04} 
so that the lattice can be viewed as ``disordered'' rather than merely defective. In the latter 
case time-independent random hopping rates, drawn from a probability distribution 
$P(p)$, are assigned to each lattice site. From now onwards we refer to this class of models 
as random-site TASEP (RST). (B) In the second class of models with quenched 
defect, the randomness is associated with the particles: a single ``impurity'' particle is 
allowed to hop at a rate $p' < p$ whereas the hopping rate of all the other particles 
is $p$. The growth of platoons of particles behind the impurity is reminiscent of 
coarsening phenomena and the steady state of the system can be expressed formally 
in terms of a Bose-Einstein-like condensate of the vacancies (absence of particles) in 
front of the impurity particle \cite{krug96,evans96,ktitarev97}. In the more general version 
of this model the particles are assigned time-independent hopping rates randomly drawn 
from a probability distribution \cite{krug96,evans96,ktitarev97,krug00}. 
From now onwards we refer to this class of models as random-particle TASEP (RPT).

In this paper we introduce a biologically-motivated quasi-one dimensional TASEP 
with two parallel tracks of lattice sites. Although the particles are not allowed to shift from 
one track to the other, the flow in each influences that in the other through a prescription 
that we define in the next section. We shall refer to this model as TASEP with mobile 
bottlenecks (TMB). As we'll explain in the next section, in one special limit this TMB 
reduces to RST. We also indicate possible extensions of the model for potential applications 
in traffic-like collective phenomena in biological systems \cite{,chou11,chowdhury13}.
Using a combination of approximate analytical arguments and highly 
accurate numerical simulations, we demonstrate the rich varieties of spatio-temporal 
organizations, including phase segregations, in this model. 

\section{The model and its comparison with other models of disordered TASEP}
\label{sec-model}

\begin{figure}
\begin{center}    
    \includegraphics[width=0.9\columnwidth]{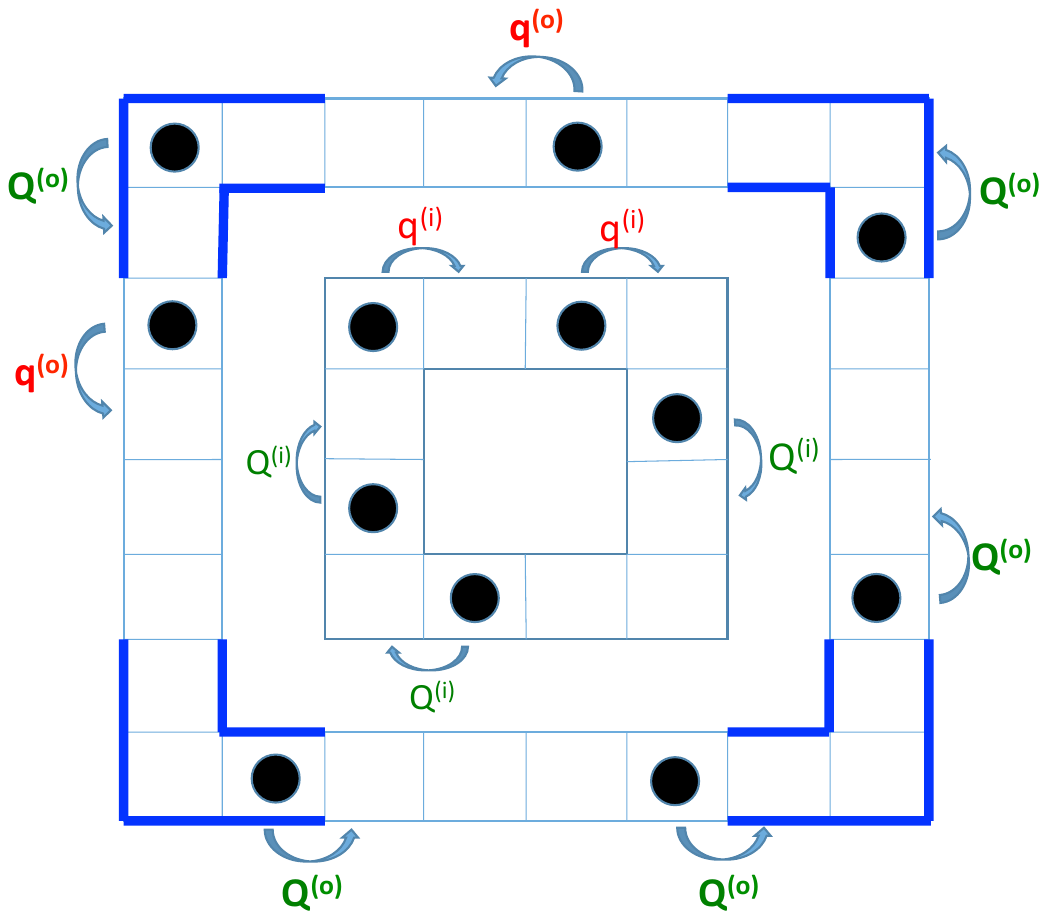}
\end{center} 
    \caption{(Color online) A schematic description of our model. The particles, represented by the black discs, on the outer lane hop unidirectionally with the probabilities $Q^{(o)}$ and $q^{(o)}$ per unit time in the absence and presence, respectively, of any hindering particle on the inner track. Similarly, the particles, also represented by the black discs, on the inner lane hop with the probabilities $Q^{(i)}$ and $q^{(i)}$ per unit time in the absence and presence, respectively, of any hindering particle on the outer track. No more than one particle can occupy the same site simultaneously. In one version the particles in both the tracks move counterclockwise whereas in the other version only the particles in the outer track move counter-clockwise while those in the inner track move clockwise (as shown in this figure).}
\label{fig-model}
\end{figure}

Dynamic blockage against directed (albeit stochastic) movement of proteins and macromolecular complexes 
along filamentous tracks is well known \cite{chowdhury13}. For example, a class of molecular motors walk along the 
surface of stiff tubular filaments called microtubules \cite{chowdhury13,kolomeisky13}. Microtubule-associated proteins, which act as blockage against the forward stepping of these motors, can detach from the microtubule thereby opening up the blockage. Similarly, several different types of molecular motors that walk along DNA or RNA strands face hindrance caused by proteins bound to the respective tracks \cite{chowdhury13}. Often these bound proteins either slide along 
the same track diffusively or, occasionally, detach from the track itself. Thus, most of the blockages against the 
movement of molecular motors are dynamic in nature. Our generic model here has been motivated by molecular motor traffic on DNA strands \cite{helmrich13} which began receiving serious attention after Brewer \cite{brewer88} summarized the earlier experimental observations scattered in the literature. During DNA replication, the replication machine, called 
DNA polymerase (DNAP) moves along one of the two strands of a double-stranded DNA that serves as its track. 
During the same period another class of machines, called RNA polymerase (RNAP) transcribes the DNA. Often 
a DNAP and a RNAP approach each other head-on along the two strands of the double stranded DNA. Because 
of the close proximity of the two tracks, each acts as a dynamic blockage for the other. The model developed 
here is motivated by such traffic-like collective movement of particles on two parallel tracks (see fig.\ref{fig-model}).

\begin{figure}
\begin{center}    
    \includegraphics[width=0.9\columnwidth]{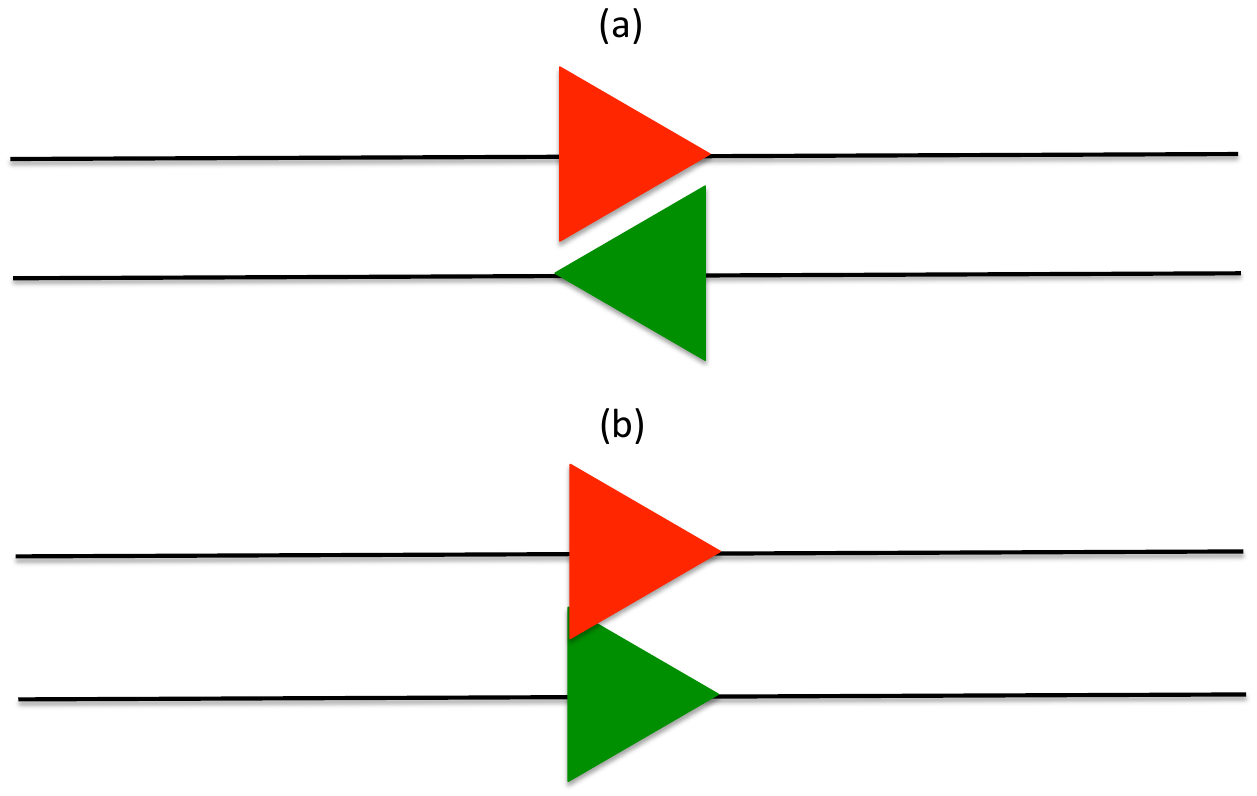}
\end{center} 
    \caption{(Color online) A schematic representation of the (a) head-on and (b) co-directional encounter of two polymerase motors moving along the two strands of a double-stranded DNA. }
\label{fig-BIOmotiv}
\end{figure} 

Our model is shown schematically in Fig.\ref{fig-model}. It consists of two square tracks with a common center where the outer and inner tracks are labelled by the symbols $T_o$ (outer) and $T_i$ (inner), respectively. Each track consists of equi-spaced discrete sites; the lengths of the outer and inner tracks, denoted by $L_o$ and $L_i$, respectively, are the total number of sites of the respective tracks. Particles  on both tracks have identical size and each covers $r$ consecutive lattice sites simultaneously (For $r>1$, it might be more appropriate to regard these particles as hard rods). 
The number of particles on $T_i$ and $T_o$ are $N_i$ and $N_o$, respectively; the corresponding {\it number densities}  being $N_i/L_i=\rho_i$ and $N_o/L_o=\rho_o$, respectively. We also define the {\it coverage densities} 
$\rho_i^{cov}=r~N_i/L_i$ and $\rho_o^{cov}=r~N_o/L_o$; for $r=1$, coverage densities are identical to the corresponding particle densities.

\begin{figure}
(i) \\
\includegraphics[width=0.5\textwidth]{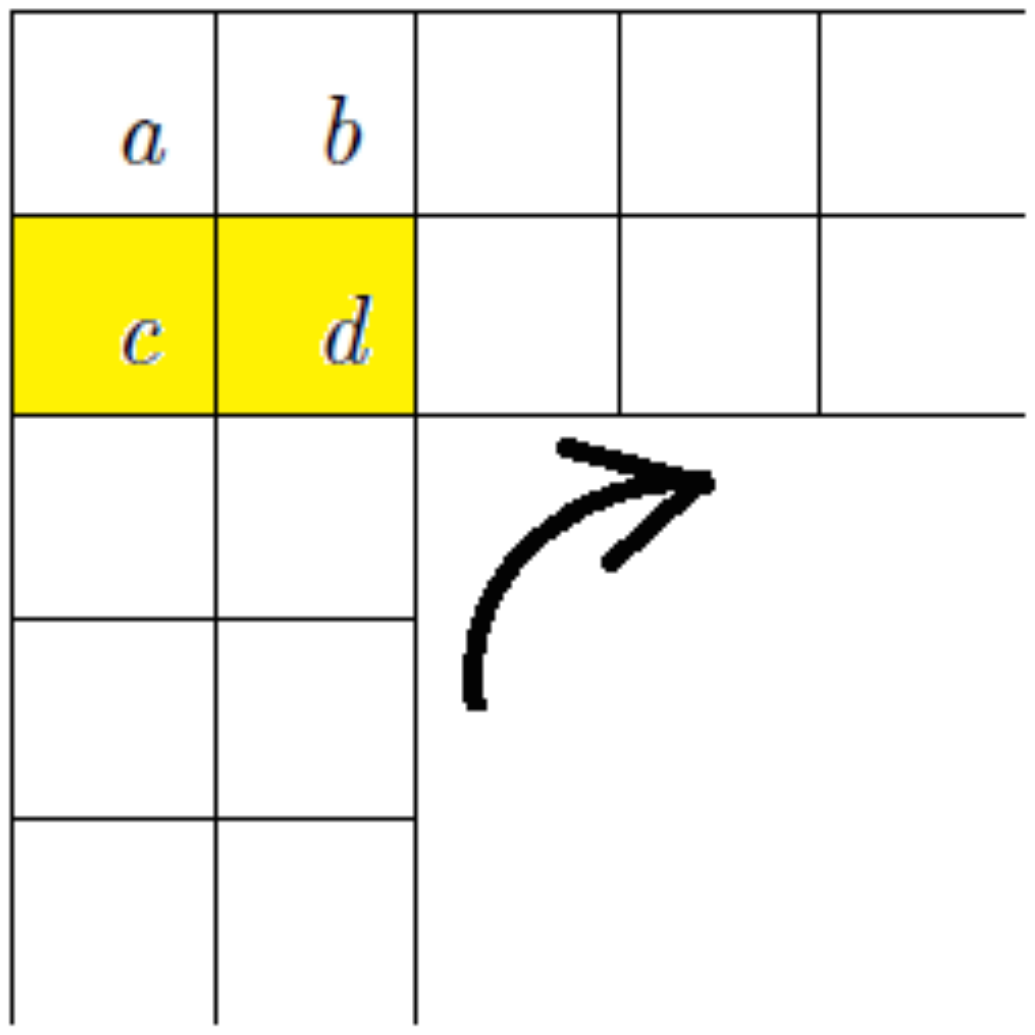} \\
(ii) \\
\includegraphics[width=0.5\textwidth]{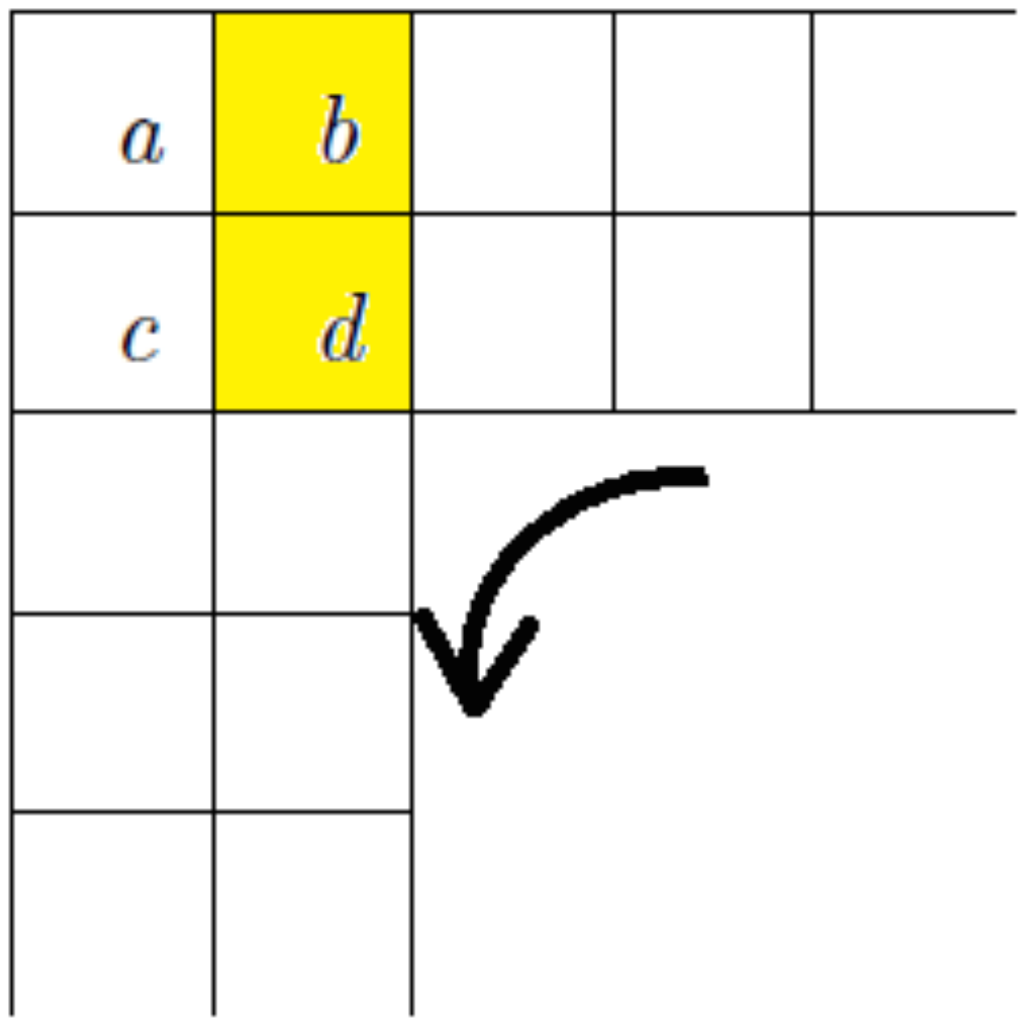}
\caption{Both (i) and (ii) in this figure depict only a small section of the system close to a corner. In (i) the particle at $d$ on the inner track, while moving clockwise, finds $c$ at a  location that is continuation of its counterparts on the outer track even before it would ``see'' $b$. Therefore, in (i) the particle at $d$ interacts with that at $c$, and not with $b$, according to the rules of the model; the particle at $b$ on the outer track remains unaffected by that at $d$. In contrast, in (ii) the particle reaches $d$ during its counter-clockwise movement on the inner track and ``sees'' $b$ before it could see $c$. Therefore, in this case, it treats $b$ as its counterpart on the outer track with which it interacts accordingly; the particle at $c$ remains unaffected by that at $d$. }
\label{fig-corners}
\end{figure}

Just as in TASEP the particles hop forward by one single lattice spacing (i.e., from one lattice site to the next on the same track) provided the target site is not covered by the leading particle. In the outer and inner tracks the hopping rates are $Q^{(o)}$ and $Q^{(i)}$, respectively provided the adjacent site on the other track is not covered by a particle; however, if the adjacent site on the other track is covered by a particle, the hopping is possible at the corresponding reduced rates $q^{(o)}$ and $q^{(i)}$, respectively. Thus, as we mentioned in the introduction, flow in the two tracks are affected by the mutual hindrance although no direct transfer of particles from one track to the other is allowed. We have studied both co-directional traffic and counter-moving traffic in the two tracks. 
{\bf 
The figure \ref{fig-corners} and explanation given in the figure removes any possible ambiguity in the definition of adjacency near the corners. 
}

Our model is very similar to the two-channel TASEP model studied by 
Popkov and coworkers \cite{popkov01,popkov03}. Further extension  
of the model by modifying the boundary conditions \cite{melbinger11} 
and coupling of one-lane TASEP with a diffusive lane \cite{evans11} 
have also been reported.  However, in those 
two-channel TASEP models the lengths of the two channels were kept 
equal. In contrast,  the emphasis here is on the effects of varying 
the ratio $f = L_{i}/L_{o}$ on the spatio-temporal organization of 
the particles. In fact, identifying the neighbor pairs on the two 
tracks, while varying $f$, is more straightforward in case of coaxial 
square tracks than in case of coaxial circular tracks.

\section{Results and discussion} 

In the following subsections, where ever possible, we provide theoretical estimates of the quantities of our interest based on either mean-field approximation or heuristic arguments. We also compute these quantities numerically by carrying out computer simulations. In order to ensure that the data are collected, indeed, in the steady state of the system we 
monitored the fluxes in both the tracks in our preliminary simulations. 
We found that, for all the system sizes and densities of our interest, the 
steady state is attained long before five million time steps. Therefore, 
for the computation of steady-state properties we 
discard the data for the first {\it five million} time steps. Throughout this paper we consider the limits $L_o \to \infty$. So far as the inner track is concerned, we'll pay special attention to the two limits $L_i \to 0$ and  $L_i \to \infty$, in addition to more general cases. Our primary interest will be the flux $J_o$ in the outer track although we'll also present the results for the inner track. The three main characteristic parameters of the inner track that control the nature of the flux $J_{o}$ in the outer track are (i) the length $L_{i}$, (ii) the density $\rho_{i}$, and (iii) $q^{(o)}$ which is a measure of the strength of the hindrance created by the inner-track particles for the outer-track flow (and, vice-versa).

\subsection{$L_{i}$ comparable to $L_{o}$} 

In order to test the validity and accuracy of the approximate analytical expressions that we derive here for $J_{o}$ and $J_{i}$, we also collect corresponding numerical data by direct computer simulations of the model. The chosen  parameter values $L_o=8000$ and $L_i=7996$  not only ensure that the tracks are sufficiently long but also allow minimal difference in the lengths of the two tracks preserving their square shapes. $P^{(i)}_j$ denotes the probability of finding a particle at site $j$ on $T_i$ (the inner track). Similarly $P^{(o)}_m$ is the probability of finding a  particle at site $m$ on $T_o$ (the outer track).

First we consider the simpler case of $r=1$. In the mean-field approximation, the master equations for the probabilities $P^{(o)}_j$ and $P^{(i)}_j$ are given by 
\begin{eqnarray}
\frac{dP^{(o)}_j}{dt}&=&[Q^{(o)}(1-P^{(i)}_{j-1})+q^{(o)}P^{(i)}_{j-1}]~P^{(o)}_{j-1}(1-P^{(o)}_{j}) \nonumber \\
                              &-& [Q^{(o)}(1-P^{(i)}_{j})+q^{(o)}P^{(i)}_{j}]P^{(o)}_{j}(1-P^{(o)}_{j+1}). \nonumber\\ 
\label{eq-masterPo}
\end{eqnarray}
\begin{eqnarray}
\frac{dP^{(i)}_j}{dt}&=&[Q^{(i)}(1-P^{(o)}_{j-1})+q^{(i)}P^{(o)}_{j-1}]P^{(i)}_{j-1}(1-P^{(i)}_{j}) \nonumber \\
                              &-& [Q^{(i)}(1-P^{(o)}_{j})+q^{(i)}P^{(o)}_{j}]P^{(i)}_{j}(1-P^{(i)}_{j+1}). \nonumber \\
\label{eq-masterPi}
\end{eqnarray} 
Consequently, in the steady state the corresponding fluxes are given by 
\begin{eqnarray} 
J_{o} = \omega^{(o)}_{mf}~ \rho_{o}~(1-\rho_{o}), ~{\rm and}~
J_{i}  = \omega^{(i)}_{mf}~ \rho_{i}~(1-\rho_{i}) \nonumber \\
\label{eq-JoutJin}
\end{eqnarray}
where the effective hopping rates under naive mean-field approximation would be 
\begin{eqnarray} 
\omega^{(o)}_{mf}&=&Q^{(o)} (1-\rho_{i})+q^{(o)}~\rho_{i}, \nonumber \\
\omega^{(i)}_{mf} &=&Q^{(i)} (1-\rho_{o})+q^{(i)}~\rho_{o} 
\end{eqnarray} 
A comparison of the mean-field predictions (\ref{eq-JoutJin}) with the corresponding simulation data revealed 
that the mean-field argument presented above leads to siginificant overestimate of both the fluxes $J_{o}$ 
and $J_{i}$.

\begin{figure}
\includegraphics[width=0.5\textwidth]{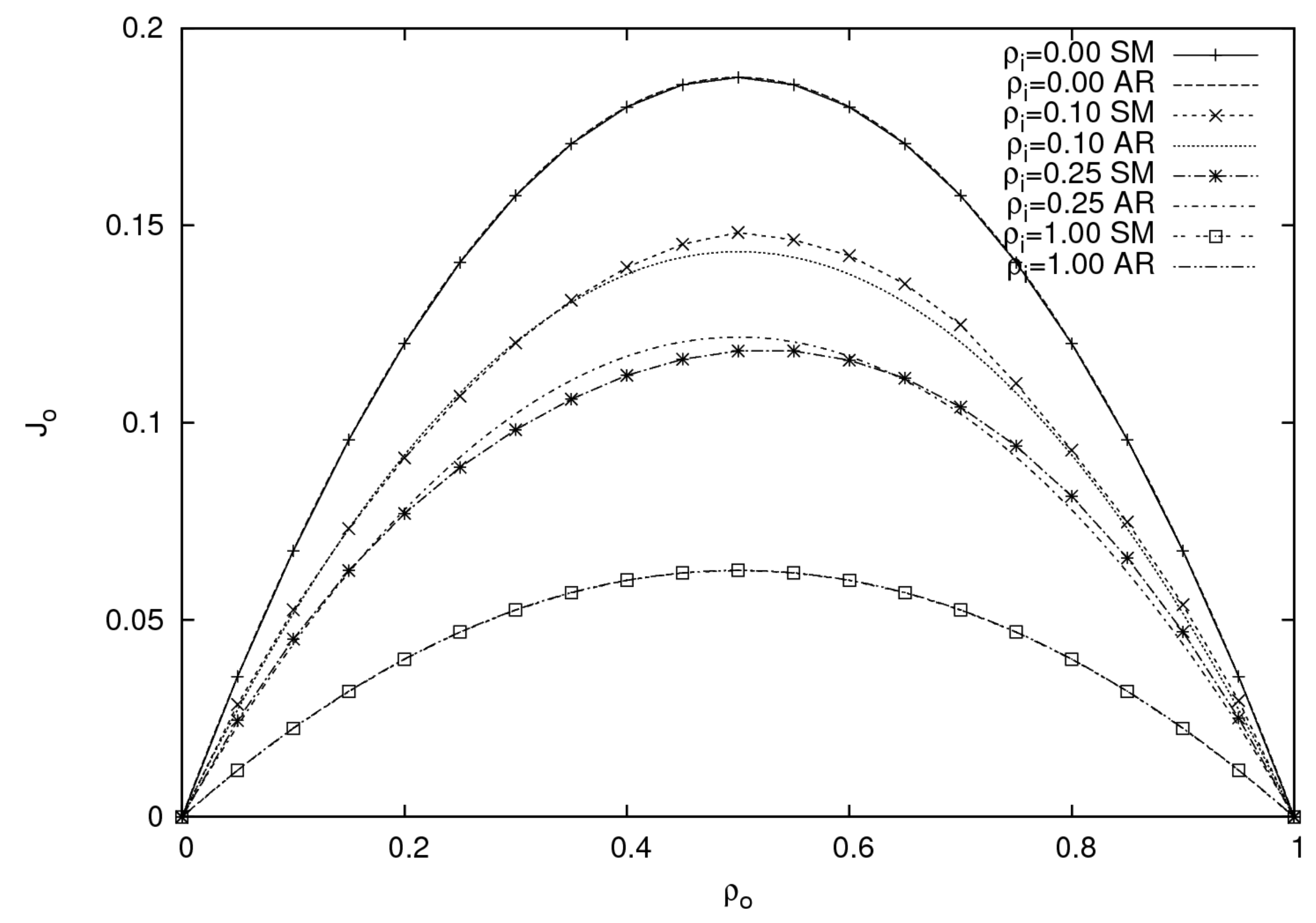}
\includegraphics[width=0.5\textwidth]{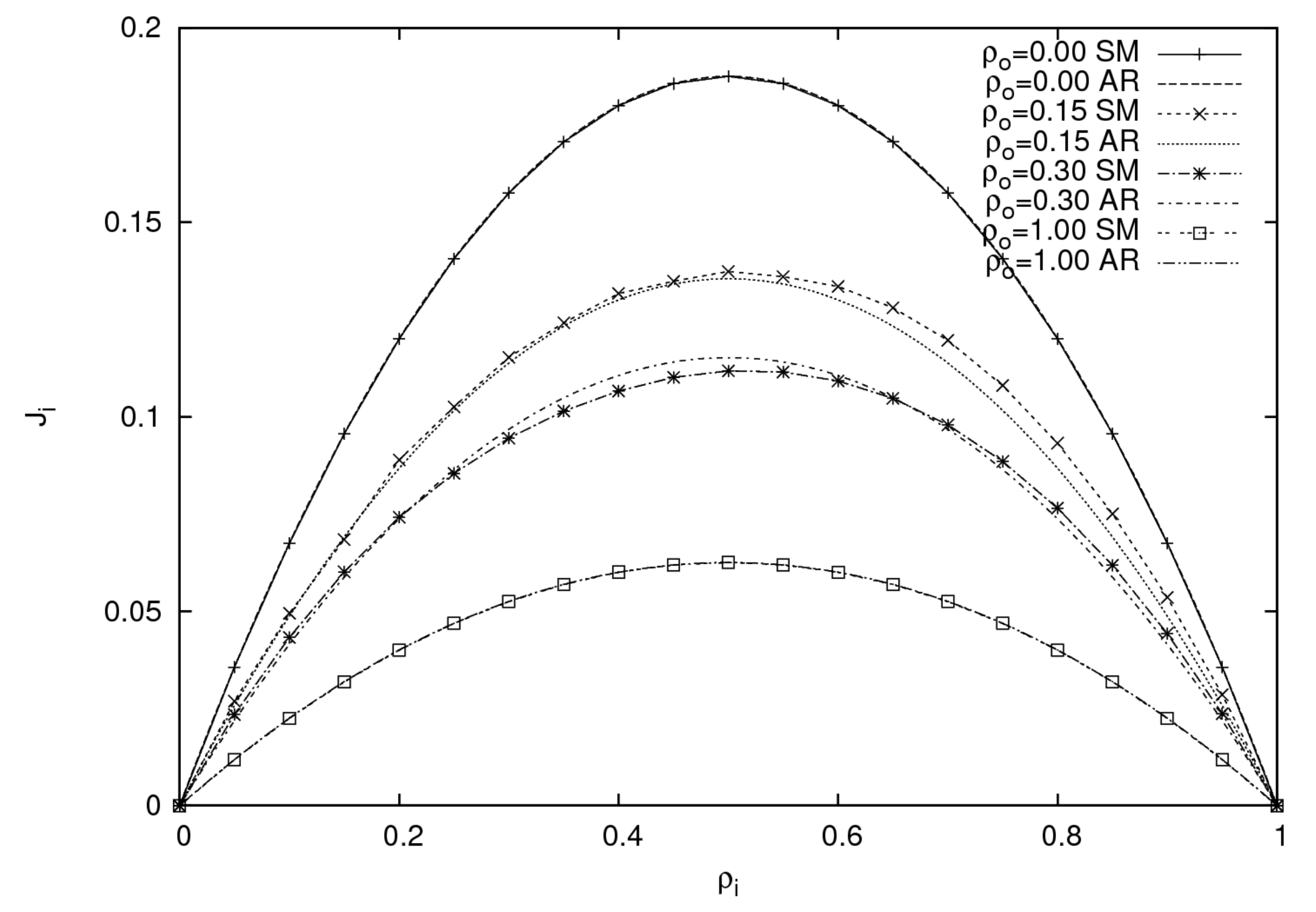}
\caption{(a) $J_{o}$ is plotted against $\rho_o$ for a few values of $\rho_{i}$. (b)  $J_{i}$ is plotted against the $\rho_i$ for a few values of $\rho_{o}$. The discrete data points have been obtained from computer simulations of the model with parameter values $L_{o}=8000$, $L_{i}=7996$, 
{\bf 
$Q^{(o)}=Q^{(i)}=0.75$ and $q^{(o)}=q^{(i)}=0.25$.
} 
The continuous curves are obtained from the 
SBA outlined above with the value of $\mu$ adjusted so as to get the best fit with the simulation data.}
\label{fig-req1FD}
\end{figure} 

\begin{figure}
\includegraphics[width=0.5\textwidth]{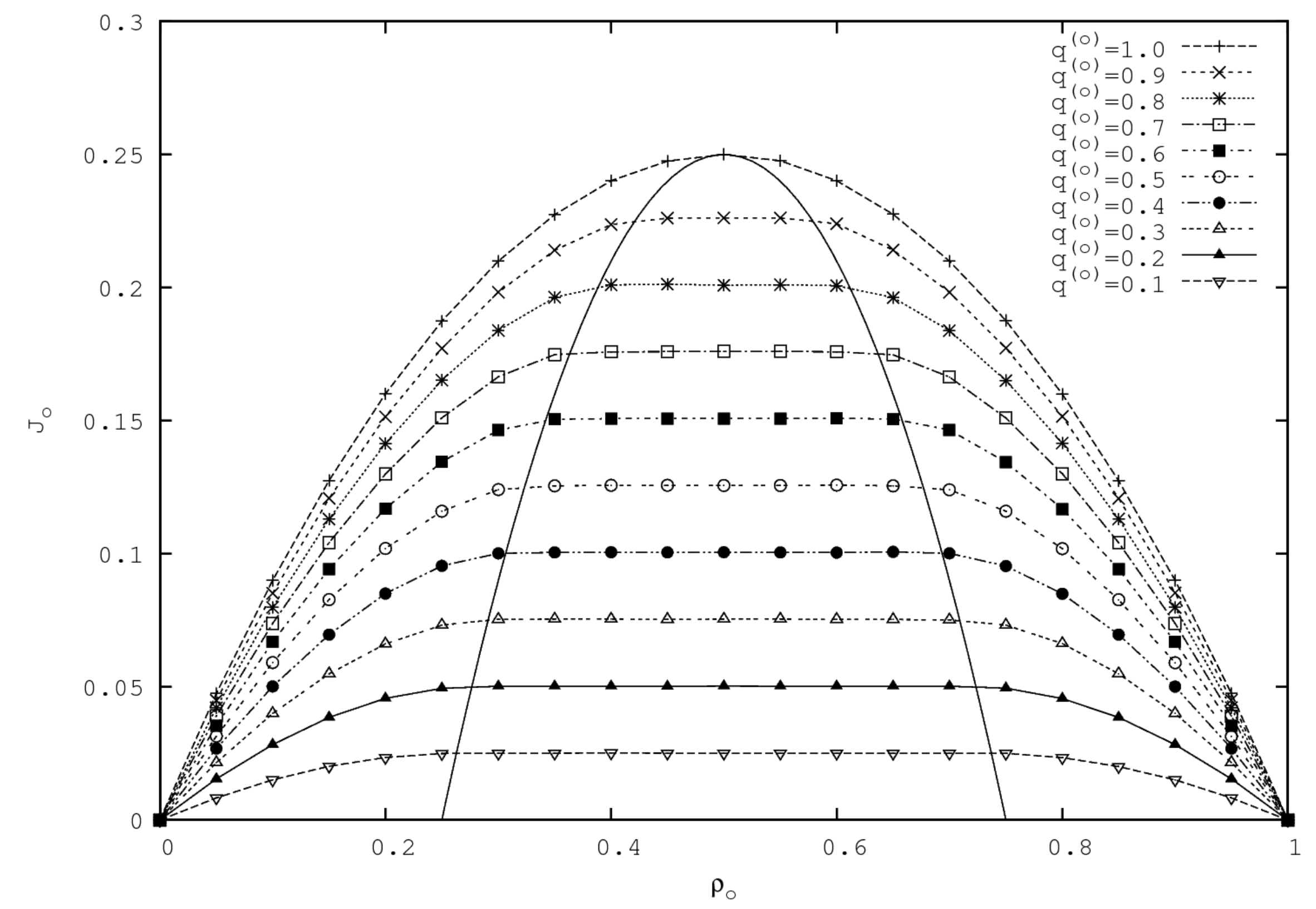}
\caption{{\bf 
$J_{o}$ is plotted against $\rho_{o}$ for $\rho_{i} = 1.0$; 
different curves correspond to different values of $q^{o}$. 
The discrete data points have been obtained from computer simulations 
of the model with parameter values $L_{o}=8000$, $L_{i}=4000$, 
$Q^{(o)}=1.0$.
}
}
\label{fig-new}
\end{figure}

In order to demonstrate what happens when the ration $f=L_{i}/L_{o}$ 
is neither approaching zero nor approaching unity, we have plotted 
$J_{o}$ against $\rho_{o}$ in fig.\ref{fig-new} for $L_{o}=8000$, 
$L_{i}=4000$. Since these data were generated keeping $\rho_{i}=1.0$, 
the innser track essentially acted as an extended bottleneck against 
the movement of the particles in the outer track. The region under 
the parabola 
\begin{equation}
J_{c} = \frac{1}{4} - (1-2\rho_{c})^2
\label{eq-parabola}
\end{equation}
drawn in fig.\ref{fig-new} corresponds to the phase coexistence region. 
This observation is fully consistent with the theory developed earlier 
in ref.\cite{tripathy97} for track-associated quenched disorder.

\begin{figure}
\includegraphics[width=0.5\textwidth]{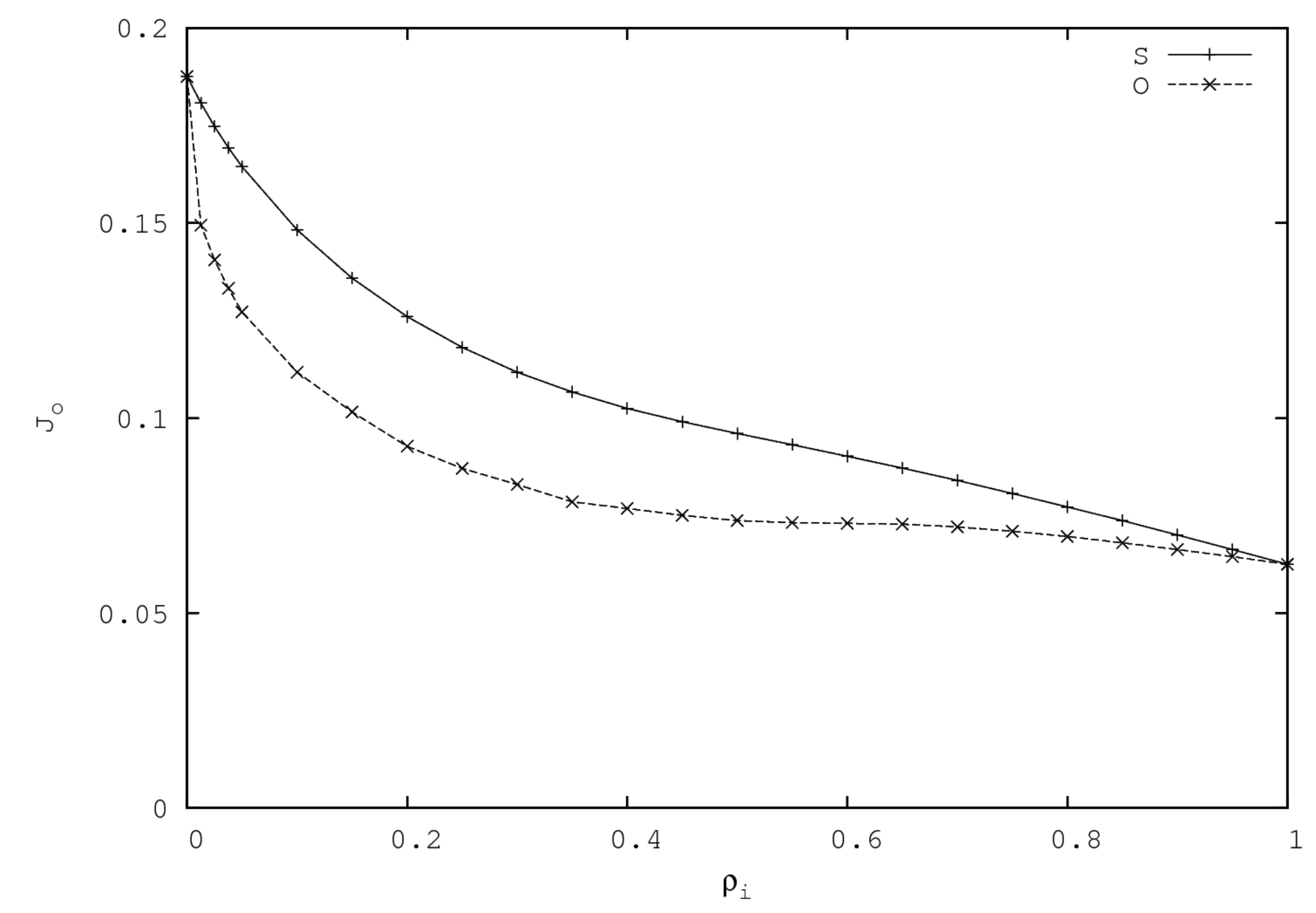}
\caption{Dependence of the flux on the relative direction of hopping of the particles in the two tracks is shown by plotting $J_{o}$ against $\rho_{i}$ for fixed $\rho_{o}=0.5$. The letters `S' and `O' correspond to the same direction and opposite direction of flow in the two tracks. The hopping rates are $Q^{(o)}=0.75=Q^{(i)}$ and $q^{(o)}=0.25=q^{(i)}$ and the lengths of the tracks are $L_{o}=8000$, $L_{i}=7996$. }
\label{fig-comp}
\end{figure}

Our mean-field estimate could be improved by extending the concept of single-bottleneck approximation (SBA) 
\cite{krug00,greulich08} that was developed for randomly distributed {\it static} defects on the track. If ${\ell}$ 
consecutive lattice sites are occupied by static defects and have defect-free sites at the two neighboring sites 
at the two ends of these ${\ell}$ sites, the ${\ell}$-site ``cluster'' acts a single bottleneck of length ${\ell}$. From 
the distribution of the lengths of such bottlenecks created by the independently distributed random defects 
one can calculate the size $<{\ell}>$ of the longest bottleneck. The effect of the defects on the flux is 
overwhelmingly dominated by the longest bottleneck. For the purpose of our calculation we replace the probability 
$\phi$ of the presence of a static defect at a lattice site by that of a mobile particle on the neighboring track. More 
specifically, for the calculation of $J_{o}$ we treat $\rho_{i}$ as the counterpart of $\phi$. Similarly, for the 
calculation of $J_{i}$ we replace $\phi$ by $\rho_{o}$. Based on these heuristic arguments the fluxes $J_{o}$ 
and $J_{i}$ are given by expressions that are identical to those in (\ref{eq-JoutJin}) except that the effective 
hopping rates $\omega^{(o)}_{mf}$ and $\omega^{(i)}_{mf}$ are replaced by the expressions 
\begin{eqnarray} 
\omega^{(o)}_{SBA} &=& Q^{(o)} e^{-\mu<{\ell}_i>}+q^{(o)}(1-e^{-\mu<{\ell}_i>})\nonumber \\
\omega^{(i)}_{SBA} &=& Q^{(i)} e^{-\mu<{\ell}_o>}+q^{(i)}(1-e^{-\mu<{\ell}_o>})
\label{eq-exponen}
\end{eqnarray}
where, utilizing the SBA \cite{greulich08}, we have
\begin{eqnarray}
 <{\ell}_i>&=&\frac{ln~ L_i+ln~(1-\rho_i)+\gamma_e}{ln~(1/\rho_i)}-\frac{1}{2} \nonumber \\
 <{\ell}_o>&=&\frac{ln~L_o+ln~(1-\rho_o)+\gamma_e}{ln~(1/\rho_o)}-\frac{1}{2}
\end{eqnarray}
Note that $\mu$ is a free parameter that is varied to get the best fit with the simulation data. 
The best fit to the simulation data are shown in fig.\ref{fig-req1FD}. It is worth pointing out that 
in spite of the increase of $<{\ell}_i>$ and $<{\ell}_o>$ with $L_i$ and $L_o$, respectively, 
the exponentials in equations (\ref{eq-exponen}) do not vanish in the thermodynamic limit 
because, as we verified, $\mu$ decreases with increasing lengths of the tracks.

The magnitude of the flux $J_{o}$ depends on the relative direction of motion of the particles in the two 
tracks. For a fixed set of parameter values, we have carried out computer simulations of the model in two different situations: (a) when particles move in the same direction in the two tracks (corresponding data are labelled by the letter `S' in fig.\ref{fig-comp}), and (b) when the particles move in opposite directions in the two tracks (corresponding data are labelled by the letter `O' in fig.\ref{fig-comp}).  $J_{o}$ is higher in case of co-directional motion than that for counter-directional motion on the two tracks. This difference is caused by the phase segregation of the particles into high-density and low-density regions. In case of opposite movements on the two tracks, a particle on the outer track has to overcome the full length of a high-density region on the inner track whereas the same particle need not face the full stretch of the same region if it moves co-directionally with the inner-track particles.

\subsection{$L_{i}$ vanishingly small compared to $L_{o}$} 

Since $L_{o} \to \infty$ and, since in this section we are assuming $L_{i} \to 4$ (the smallest allowed value of 
$T_{i}$ that preserves its shape is $4$), the inner track effectively acts as a ``point'' defect for the particles 
moving on the outer track.

\begin{figure}
 \includegraphics[width=0.5\textwidth]{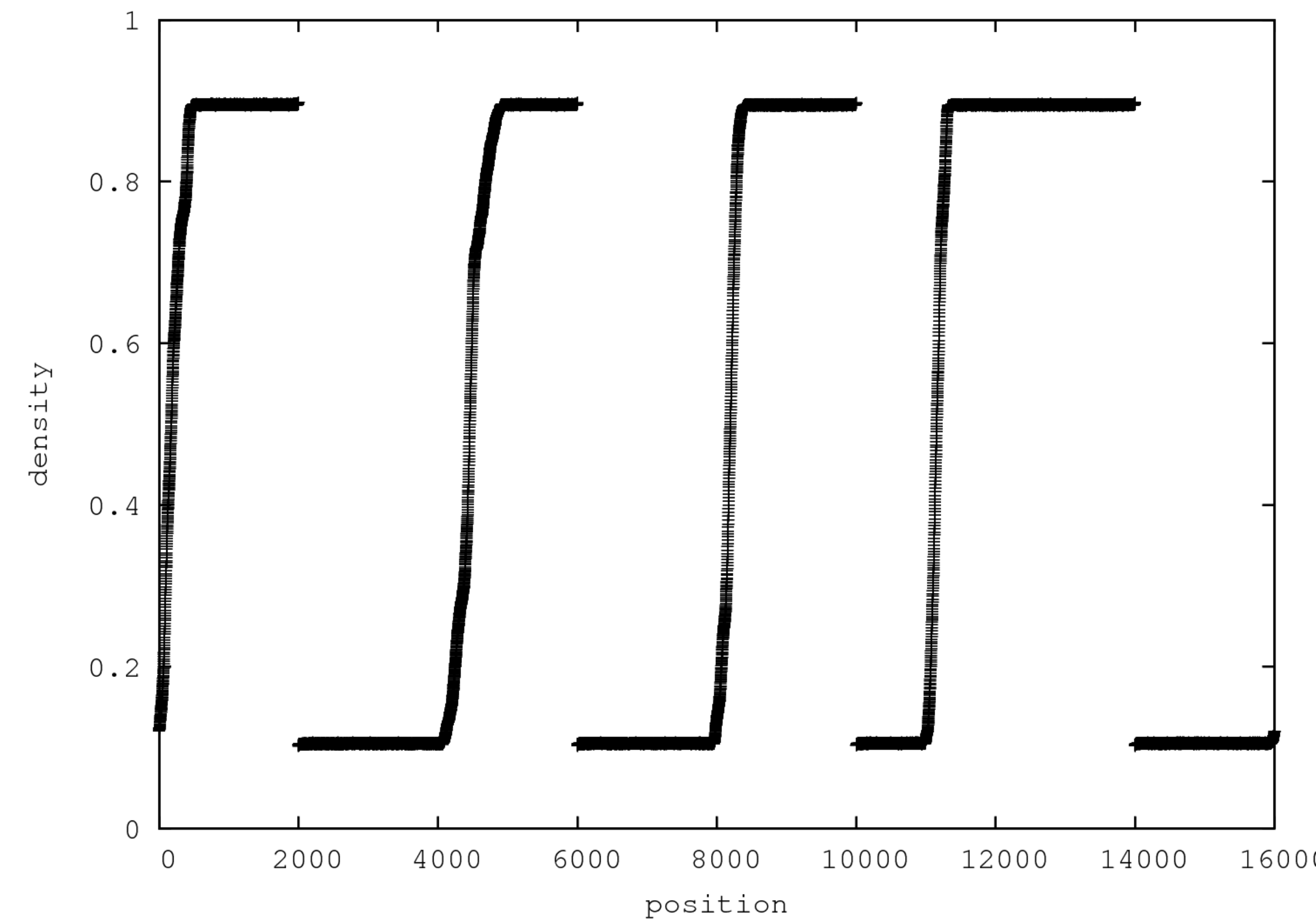}
 \includegraphics[width=0.5\textwidth]{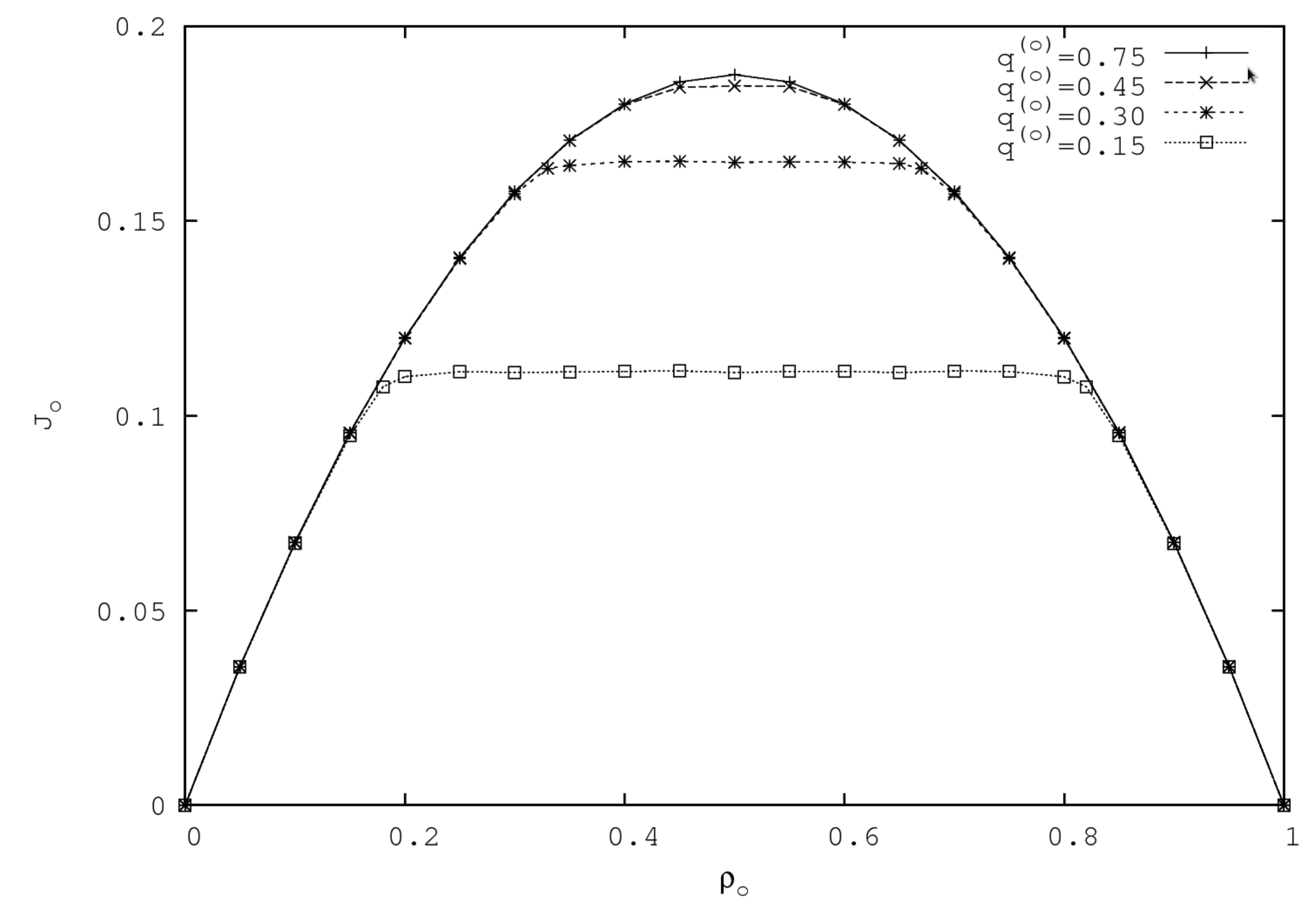}
\caption{(a) The density profile on the outer track for $\rho_{i}=1.0$, $\rho_{o}=0.5$, $Q^{(o)}=0.9$, $q^{(o)}=0.1$.
(b) $J_{o}$ is plotted against $\rho_{o}$ at $\rho_i=1.0$ for $Q^{(o)}=0.75$ and a few different values of the parameter $q^{(o)}$. For both (a) and (b) the lengths of the tracks are $L_{o}=16000$, $L_{i}=4$.}
\label{fig-rhoi1PhCo}
 \end{figure}

\begin{figure}
 \includegraphics[width=0.5\textwidth]{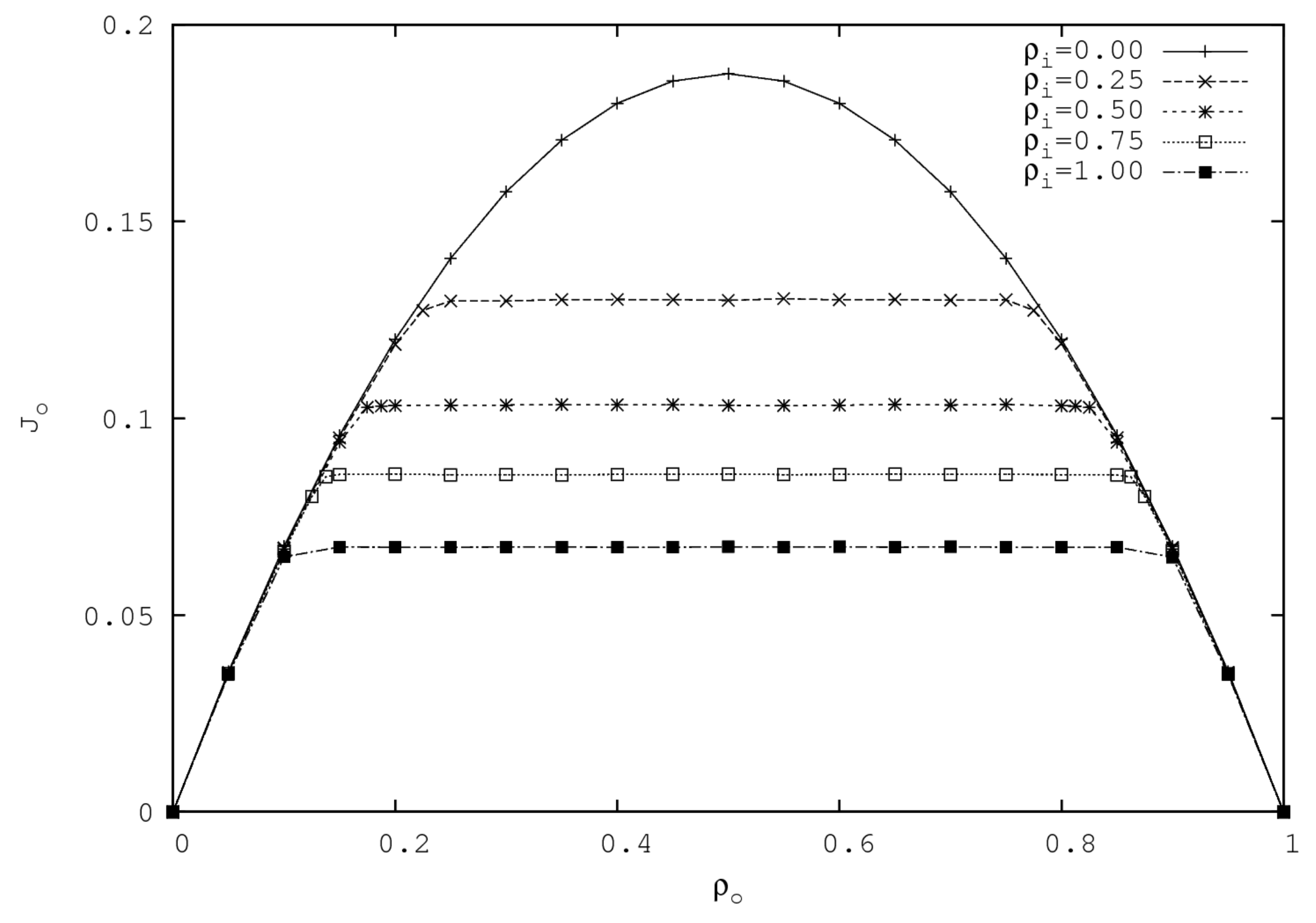}
\caption{$J_{o}$ is plotted against $\rho_{o}$ for several different values of $\rho_i$. The values of the other parameters used for this figure are $L_{o}=8000, L_{i}=80, Q^{(o)}=0.75=Q^{(i)}, q^{(o)}=0.25=q^{(i)}$.}
\label{fig-rhoi5PhCo}
 \end{figure}

First let us consider the special case $\rho_{i}=1$. In this case the inner track acts effectively as a static ``point-like'' defect. It is well known that macroscopic phase segregation takes place in a TASEP with a single point defect, where the density profile exhibits coexistence of a high-density and low-density regimes. Since each of the four sides of the outer track in our model sees a point-like defect, the profile consists of four segments each of which is similar to that of a TASEP with a single point-defect (see fig.\ref{fig-rhoi1PhCo}). The flat symmetric plateaus observed on the plots of $J_{o}$ against $\rho_{o}$ correspond to the coexistence of the high-density and low-density regimes. This qualitatively similar to the corresponding flux-density diagrams for TASEPs with point defects. Moreover, as shown also in the fig.\ref{fig-rhoi1PhCo}, lower is the value of $q^{(o)}$, the stronger is the effect of the bottleneck and the smaller is the flux $J_{o}$.

Next, in order to demonstrate the more general situations, in fig.\ref{fig-rhoi5PhCo} we have plotted $J_{o}$ as a function of $\rho_{o}$ for five values of the parameter $\rho_{i}=0.0, 0.25,0.5.0.75,1.0$. As the data in this figure show, the higher is the density $\rho_{i}$, loger is longest bottleneck and, consequently, the wider is the plateau region. This is consistent with the fact that a single bottleneck of longer size lowers the flux to a small value \cite{chowdhury00}. 
Finally, as expected intuitively, the flux-density curve for the outer track approaches the parabolic form $Q^{o} \rho_{o}(1-\rho_{o})$ in the limit $\rho_{i} \to 0$.

\begin{figure}
\includegraphics[width=0.5\textwidth]{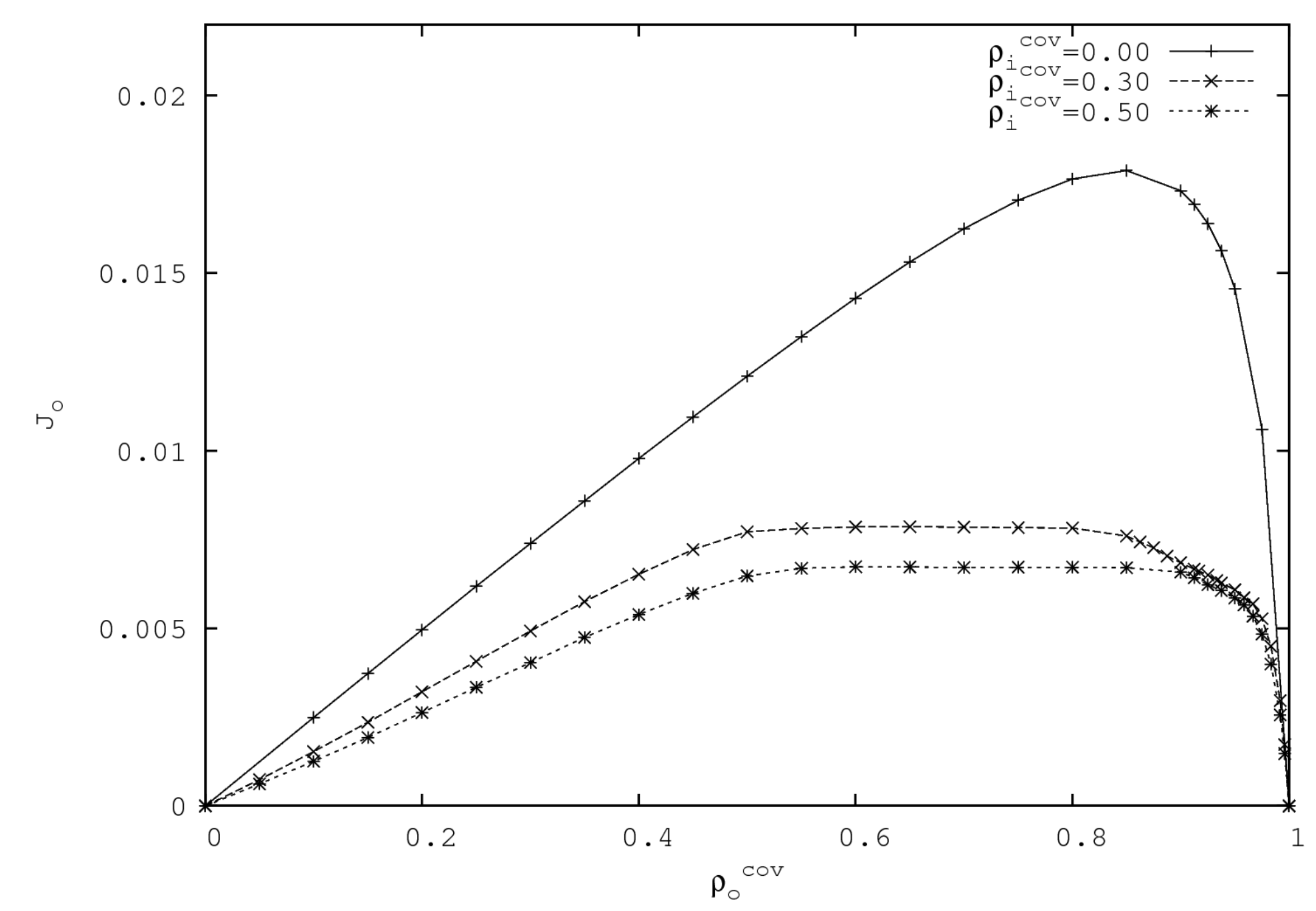}
\caption{$J_o$ of particles with $r=30$ is plotted against $\rho_{o}^{\rm cov}$ for three different values of  $\rho_{i}^{\rm cov}$; the hopping rates are $Q^{(o)}=0.75=Q^{(i)}$ and $q^{(o)}=0.25=q^{(i)}$.}
\label{fig-rgt1FD}
\end{figure}

\subsection{Flux-density relation for particles of size $r > 1$}

Keeping in mind the possible future application of extensions of our model to the biophysical phenomena mentioned in the introduction, we have also studied the model for particle size $r$ larger than $1$. As a typical example, the flux $J_{o}$ is plotted against the corresponding coverage density $\rho_{o}^{\rm cov}$ in fig.\ref{fig-rgt1FD} in those situations where all the particles on both the tracks have the same length $r=30$. The plateau observed in the special case $r=1$ survives also for all $r>1$. However, as is well known for TASEP with hard rods \cite{chowdhury00,schadschneider10}, the maximum of the flux shifts to increasingly higher densities with increasing $r$.

\section{Summary and Conclusions}

In this paper we have developed a {\bf simple} model for the TASEP on two co-axial square tracks that influence each other without any transfer of particles from one to another. In spite of its extreme simplicity, the model exhibits rich variety of phenomena in different parameter regimes. In the limit $L_{i}/L_{o} \to 0$, the model reproduces the known properties of a TASEP with a single ``point-like'' defect on the track. In particular, it exhibits macroscopic phase segregation of the particles on the outer track. In general, our model the width of the coexistence region can be tuned by varying three distinct control parameters which we have clearly identified. Extensions of this model, incorporating further details, are likely to find use in modeling the traffic-like collective biological phenomena mentioned in section \ref{sec-model}.

\section*{ACKNOWLEDGEMENTS}

This work is supported by a KVPY Fellowship (SS) and a J.C. Bose National Fellowship (DC).



\end{document}